\documentclass[graybox]{svmult}

\usepackage{type1cm}        
\usepackage{makeidx}         
\usepackage{graphicx}        
\usepackage{multicol}        
\usepackage[bottom]{footmisc}

\makeatletter
\newcommand*{\rom}[1]{\expandafter\@slowromancap\romannumeral #1@}
\makeatother

\usepackage{newtxtext}       %
\usepackage{newtxmath}       

\usepackage[square,numbers]{natbib}

\usepackage{array}

\makeindex             

\begin{document}

\title*{Decoding the deterministic nature of black hole IGR J17091–3624}
\author{Anindya Guria and Banibrata Mukhopadhyay}
\institute{Anindya Guria \at Indian Institute of Science, CV Raman Road 560012, \email{anindyaguria@iisc.ac.in}
\and Banibrata Mukhopadhyay \at Indian Institute of Science, CV Raman Road 560012, \email{bm@iisc.ac.in}}
%
%
\maketitle
\vspace{-2cm}
\textit{To be published in Astrophysics and Space Science Proceedings, titled "The Relativistic Universe: From Classical to Quantum, Proceedings of the International Symposium on Recent Developments in Relativistic Astrophysics", Gangtok, December 11-13, 2023: to felicitate Prof. Banibrata Mukhopadhyay on his 50th Birth Anniversary", Editors: S Ghosh \& A R Rao, Springer Nature}\\

\abstract{The differentiation between chaotic and stochastic systems has long been scrutinized, particularly in observations where data is often noise-contaminated and finite. Our research examines the dual nature of the black hole X-ray binary IGR J17091–3624, an object whose behavior has been closely studied in parallel to GRS 1915+105. Remarkable similarities in the temporal classes of these two objects are explored in literature. However, this was not the case with their non-linear dynamics: GRS 1915+105 shows signs of determinism, while IGR J17091-3624 was found to be stochastic. In this study, we confront the inherent challenge of noise contamination, as in IGR J17091–3624, faced by previous studies, particularly Poisson noise, which adversely impacts the reliability of non-linear results. We employ several denoising techniques to mitigate noise effects and employ methods like Autoencoder, Principal Component Analysis (PCA), Singular Value Decomposition (SVD), and Correlation Integral (CI) to isolate the deterministic signatures. We have found signs of determinism in IGR J17091–3624 after denoising, thus supporting the hypothesis of it being similar to GRS 1915+105, even as a dynamical system.}

\abstract*{The differentiation between chaotic and stochastic systems has long been scrutinized, particularly in observational settings where data is often noise-contaminated and finite. Our research examines the dual nature of the black hole X-ray binary IGR J17091–3624, an object whose behavior has been closely studied in parallel to GRS 1915+105. Remarkable similarities in the temporal classes of these two objects are explored in literature. However, this was not the case with their non-linear dynamics: GRS 1915+105 shows signs of determinism, while IGR J17091-3624 was found to be stochastic. In this study, we confront the inherent challenge of noise contamination, as in IGR J17091–3624, faced by previous studies, particularly Poisson noise, which adversely impacts the reliability of non-linear results. We employ several denoising techniques to mitigate noise effects and employ methods like Autoencoder, Principal Component Analysis (PCA), Singular Value Decomposition (SVD), and Correlation Integral (CI) to isolate the deterministic signatures. We have found signs of determinism in IGR J17091–3624 after denoising, thus supporting the hypothesis of it being similar to GRS 1915+105, even as a dynamical system.}

\section{Introduction}
\label{sec:Intro}
GRS~1915+105, a well-studied black hole X-ray binary, exhibits a high spin and quasi-periodic oscillations (QPOs) across a wide frequency range \citep{Mirabel1994, belloni2001QPO}. It is categorized into 12 temporal classes \citep{belloni2000} and identified in three spectral states. Certain classes demonstrate deterministic, chaotic, or fractal characteristics, contrasting with others that appear stochastic (S) \citep{Adegoke2018}.

IGR J17091-3624, often considered a twin to GRS1915+105 due to shared features like``heartbeat" oscillations and high-frequency QPOs \citep{Rao_2012, Altamirano_2011}, is however fainter and exhibits lower mass and spin \citep{Rao_2012}. Nevertheless, its temporal classes were previously found to be S or noise-dominated \citep{Adegoke2020}. This calls into question the true nature of its dynamics, especially since X-ray lightcurves are heavily contaminated by Poisson noise. Previous studies suggested increasing bin size might increase fractal dimensions, though conclusive results are lacking due to insufficient data \citep{Adegoke2020}.

We plan here to revisit the problem and rigorously analyze the timeseries of the temporal classes of IGR~J17091-3624, with special considerations to address the noise. Our approach is two-pronged: first, we apply a variety of filtering techniques to reduce the effect of Poisson noise in the lightcurves. These methods include moving averages, non-local means \citep{buades2004image, Buades_NLM2005}, and an adaptive algorithm optimized for detecting determinism in heavy-noise environments \citep{Tung2011}. Second, we explore beyond the traditional correlation integral (CI) method \citep{GRASSBERGER1983189, Misra:2006qg, harikrishnan2006, Adegoke2018, Adegoke2020} to falsify the hypothesis that all IGR~J17091-3624 temporal classes are S. Some of these techniques include principal component analysis (PCA) \citep{PCA_SVD}, singular value decomposition (SVD) \citep{PCA_SVD, Misra:2006qg}, and a novel machine learning-based autoencoder \citep{Pradeep2023} that has been optimized to search for scale-invariant features in lightcurves (both time and frequency domains). Interestingly, we find that several temporal classes of IGR~J17091-3624 show promising signs of being non-stochastic (NS), hence refuting the earlier hypothesis. Through this work, we attempt to highlight the importance of utilizing filtering techniques and designing novel tests (e.g., the autoencoder, used in this paper) while performing non-linear analysis of noisy astrophysical timeseries data.


\section{Difficulties Encountered Earlier with IGR~J17091-3624} \label{sec:difficulties}
Determining whether a dynamical system is S or NS presents several challenges in experiments or observations where data are finite and typically noise-contaminated. In earlier studies of IGR J17091-3624, no NS behaviors were evident, contrasting with the well-established deterministic nature of certain temporal classes of GRS 1915+105. However, the latter exhibits about 20 times more average photon counts than IGR J17091-3624, as recorded by RXTE (PCA). This argues for a significant influence of Poisson noise due to the lower photon counts in IGR J17091-3624. Poisson noise, inherent in photon counting devices and increasing with fewer photons, complicates the interpretation of nonlinear dynamics from such data.

Previous study \cite{Adegoke2020} noted potential deviations from stochasticity with increased binning times from $0.125$ s to $0.5$ s, although the reduction in data points hampered definitive conclusions using the CI method. Nevertheless, increasing the bin size can reduce some effects of Poisson noise, suggesting that advanced noise-reduction techniques might allow a more accurate analysis of the nonlinear characteristics of these sources.

\section{Proposed Filtering Methods} \label{sec:proposed filtering}
Based on the experience from previous efforts, we can list some key properties that any filtering method should ideally satisfy. The rationale behind many of these points was discussed succinctly earlier \cite{Kostelich1993}, given below as: 
\begin{enumerate}
    \item There should be no implicit or explicit dependence on the signal's power spectrum. This is because chaotic timeseries are expected to have a broad spectral spread. Also, the power spectrum of Poisson noise of an ideal Poisson process is uniform: traditional low pass or band pass filtering would not be effective in our application.
    \item Our filtering technique should avoid introducing additional local correlation in our data. It would ensure that our conclusions are not affected by the filtering process.
    \item The number of data points should not be reduced drastically by the filtering algorithm. As already mentioned, this enables the use of the CI method.
   
\end{enumerate}
We shall begin by describing some trivial filters that satisfy our requirements and gradually move on to some of the more sophisticated ones.

\subsection{Convolution-based Filters (BOX and GAU)}

Convolution filters are foundational in Digital Signal Processing (DSP) for signal manipulation and enhancement. 
A detailed exploration of convolution kernels can be found in standard texts such as \cite{Smith_Steven_W_1997}.
Consider a normalized kernel \(g_i\), where \(\sum_{i=-k}^{k}g_i=1\), applied to an input series \(x_n\) to produce an output \(y_n\) defined by $y_n=\sum_{i=-k}^{k}g_ix_{n+i}$.
Popular kernels include boxcar (BOX) and Gaussian (GAU), each chosen for their specific advantages in various applications. In this study, we have made use of both. BOX is nothing but a moving average; although it is extremely simple ($g_i=1/(2k+1)$, $y_n=\sum_{i=-k}^{k}x_{n+i}/(2k+1)$), due to its discontinuous edges it is not ideal for preserving spectral properties. To remedy this, the GAU kernel is used ($g_i=Ce^{-\frac{i^2}{2\sigma^2}}$, $y_n=C\sum_{i=-k}^{k}e^{-\frac{i^2}{2\sigma^2}}x_{n+i}$). Due to the central limit theorem, GAU is mathematically identical to an infinite number of passes by a BOX filter. 



\subsection{Non-local Means (NLM) Denoising}
Initially devised for image denoising, the non-local means (NLM) algorithm mitigates noise by leveraging similar samples or pixels, regardless of their spatial proximity. 
A detailed discussion of the algorithm can be found in \cite{buades2004image, Buades_NLM2005}.
To adapt NLM for one-dimensional signals, we begin with a boxcar convolution to generate an auxiliary series \(B_n = \sum_{i=-k}^{k} x_{n+i}/(2k+1)\), using a kernel of size 9. The similarity between data points at indices \(m\) and \(n\) is quantified using a Gaussian weight function \(f_{mn}= \exp\left(-(B_m - B_n)^2/h^2\right)\), where \(h\) is set based on the noise standard deviation. We choose \(h\) based on the standard deviation within each boxcar smoothed ``patch." The denoised timeseries is then computed as \(y_n = \frac{1}{C_n} \sum_{m=1}^{N} f_{mn} x_m\) where \(C_n = \sum_{m=1}^{N} f_{mn}\) is a normalization factor. Although NLM may leave behind low-amplitude white noise, its impact on nonlinear analysis is minimal \cite{buades2004image}.

\subsection{Adaptive Denoising Algorithm (ADA)}
This technique is suitable for detecting chaos/determinism within heavy noise \cite{Tung2011}. 
Initially, the algorithm divides a timeseries into segments or windows of length $w = 2n + 1$ points, with $n+1$ points overlapping among them. A $k^{th}$ order polynomial is fitted for each segment. For the $i^{th}$ and $(i+1)^{th}$ segments let them be $p^k_i(l_1)$ and $p^k_{i+1}(l_2)$, where $l_1,l_2\in\{1,2,\cdots 2n+1\}$. The denoised data within this $i^{th}$ overlapped region are found using the weights $w_1=\left[1-\frac{(l-1)}{n}\right]$ and $w_2=\left[\frac{(l-1)}{n}\right]$ as
\[\begin{array}{cc}
y_{ni+l}=w_1 p^k_i(l+n)+w_2 p^k_{i+1}(l) & l\in\{1,2,\cdots n+1\}.
\end{array}\]
This choice of weighting eliminates discontinuities at the boundaries of neighboring segments; this fitting ensures continuity all across the data.
Two free parameters for fine-tuning are the segment size $(w)$ and polynomial order $(k)$. 
We have followed the recommended protocol \citep{Tung2011} to determine their optimum values.

\section{Methods for Classification of S and NS Timeseries} \label{sec:Classification}
 The CI method has been considered for quite some time for analyzing non-linearity in black hole lightcurves, e.g.  \citep{Adegoke2018,Adegoke2020,Misra:2006qg}. However, recently other techniques have also been explored and found to be viable for classifying temporal classes of GRS~1915+105. We have successfully used several of these methods in this work, notably matrix-based approaches like PCA and SVD \cite{PCA_SVD}, and a machine learning-based method using autoencoders \citep{Pradeep2023}.


\subsection{Correlation Integral Method and Surrogate Analysis} \label{subsec:CI method}
In the delay-embedding technique of CI originally proposed by Grassberger-Procaccia (GP), first vectors are constructed in an $M$-dimensional reconstructed phase space with delay $\tau$, where each vector \(\Vec{\xi_i}=[x_{i}, x_{i+\tau}, \dots, x_{i+(M-1)\tau}]\) represents the system's state at different times \citep{Fraser1986}. The embedding parameters 
are chosen based on the first minimum of the mutual-information function \citep{Adegoke2020}.

The correlation integral \(C_M(r)\) estimates the probability that two randomly selected vectors are within a distance \(r\) 
with the scaling law of $C_M(r) \sim r^{D_2}$ such that $D_2$ is the correlation dimension \citep{GRASSBERGER1983189}. The scaling of \(D_2\) linear with \(M\) suggests S behavior; an eventual saturation of \(D_2\) suggests deterministic dynamics.

Surrogate analysis tests the null hypothesis that the timeseries is generated by a linear stochastic process \citep{Misra:2006qg}. We generate a set of surrogate timeseries that matches the original data in probability distribution and power spectrum using the IAAFT method \citep{Venema_IAAFT2006}. The Normalised Mean Sigma Deviation (NMSD) measures differences in \(D_2\) determined from the original and surrogate data:
\[
nmsd^2=\frac{1}{M_{\max }-M_{\min }} \sum_{M=M_{\min }}^{M_{\max }}\left[\frac{D_2(M)-\left\langle D_2^{\mathrm{surr}}(M)\right\rangle}{\sigma_{\mathrm{SD}}^{\mathrm{sur}}(M)}\right]^2.
\]
An \(nmsd > 3\) for 19 surrogates suggests NS processes \citep{harikrishnan2006} (with a confidence of 95\%). 

\subsection{Singular Value Decomposition (SVD)}\label{subsec:SVD}
SVD is employed to analyze the temporal dynamics of timeseries data using matrix techniques. 
The first two components, \(E1\) and \(E2\), of the right singular vector describe the dominant temporal dynamics, 
visualized by plotting \(E1\) versus \(E2\), and analyzed using Betti numbers as topological descriptors \citep{Betti1870}. For a two-dimensional manifold, we consider Betti numbers, \(\beta_0\) (the number of connected components) and \(\beta_1\) (the number of 1-dimensional holes). S timeseries typically show a simple topology \((\beta_0, \beta_1) = (1, 0)\), whereas NS signals exhibit more complex topologies with multiple components and holes.

The S and NS timeseries are classified based on the sum of Betti numbers. If $\beta_0 + \beta_1 = 1$, then we call it S, else if $\beta_0 + \beta_1 > 1$ then it is NS. The number of connected components is identified using DBSCAN \citep{Martin1996} from the scikit-learn package \citep{scikit-learn}, with more complex structures assessed manually. Detailed results are presented in subsection \ref{subsec: SVD_result}.

\subsection{Autoencoder Based Approach for Measuring Deviation from Stochasticity (\(DS\))}\label{subsec:Autoencoder}
Utilizing a method proposed recently \cite{Pradeep2023}, we employ autoencoder-based techniques to detect deviations from stochasticity in timeseries. This method leverages time-invariant features across multiple timescales in frequency and time domains through multi-scale reconstruction and peak analysis in dissimilarity curves.

The process involves segmenting the original timeseries, applying Discrete Fourier Transform (DFT) on each segment, and training two autoencoders—one for the time domain and another for the frequency domain data. These autoencoders, trained on a mixture of stochastic, deterministic, and noise-contaminated signals (including GRS 1915+105 lightcurves), reconstruct features to generate dissimilarity (\(D_t\)) and prominence-of-peaks (\(P_t\)) curves. 

For multi-scale analysis, these curves are segmented into windows (\(W_k\)), and both bias-corrected dissimilarity (\(\Tilde{d^k_t}\)) and the prominence (\(p^k_t\)), are used to compute the Kullback-Leibler divergence and the coefficients of variance (\(CV1\) and \(CV2\)), respectively. \(CV1\) reflects the variance across reconstructions at different scales, while \(CV2\) assesses changes in peak prominence, with lower values expected for S signals.

The deviation from stochasticity \(DS\) is calculated as:
\[
DS = \frac{CV1 \times CV2}{100},
\]
with a critical threshold at \(DS=1.5\) to distinguish between S and NS timeseries, as determined using an SVM classifier \cite{Pradeep2023}. A value of \(DS \geq 1.5\) indicates NS behavior, while \(DS < 1.5\) suggests S characteristics. Detailed results of this approach are discussed in subsection \ref{subsec: Autoencoder results}.

\subsection{Principle Component Analysis and DBSCAN} \label{subsec: PCA}
PCA is employed for dimensionality reduction, focusing on eigenvalue ratios from the covariance matrix of timeseries data to detect dominant directions indicative of NS behavior. 
If the respective largest eigenvalues of two segments, $\lambda_1$ and $\lambda_2$, of the covariance matrix \cite{PCA_SVD} follow $\lambda_1 \geq \lambda_2$, the ratio $r=\lambda_1/\lambda_2$ signifies the presence of a dominant direction, a characteristic of NS timeseries.

The key features derived from the eigenvalue ratio (ER) include:
\begin{enumerate}
    \item \textbf{MER}: Maximum ER, lower in S.
    \item \textbf{VAR}: Variance of ER, also lower in S.
    \item \textbf{Area}: Total area under the ER curve, higher in NS.
\end{enumerate}

Each timeseries is represented as a point in 3D (MER, VAR, Area) space using these features. DBSCAN, an unsupervised clustering algorithm, is then applied to segregate S and NS signals by identifying dense clusters and outliers. The formalized use of PCA and DBSCAN resolves prior ambiguities in classifying timeseries data, with detailed results discussed in subsection \ref{subsec: PCA result}.

\section{Results}
\subsection{Correlation Integral Results} \label{subsec: CI results}
Tables \ref{tab:CI1} and \ref{tab:CI2} present $nmsd$ and $D_2$ values for unfiltered and denoised signals. We follow the criterion that $D_2$ values are reported only when $nmsd \ge 3$ \citep{harikrishnan2006}. A class is labeled as S or NS based on a consensus of at least three filtering techniques. Classes with conflicting results from the methods are labeled ``S/NS".




\begin{table}[!h]
\caption{Correlation Integral results of IGR J17091-3624 (Part 1: Original, GAU, and BOX Data)}
\label{tab:CI1}       
%
\centering
\begin{tabular}{
  >{\centering\arraybackslash}p{2cm}  
  >{\centering\arraybackslash}p{1.2cm}  
  >{\centering\arraybackslash}p{1.2cm} 
  >{\centering\arraybackslash}p{1.2cm}  
  >{\centering\arraybackslash}p{1.2cm} 
  >{\centering\arraybackslash}p{1.2cm}
  >{\centering\arraybackslash}p{1.2cm}
  >{\centering\arraybackslash}p{1.2cm}}
\hline\noalign{\smallskip}
ObsID & Class & $nmsd$ & $D_2$ & $nmsd$ & $D_2$ & $nmsd$ & $D_2$ \\
{} & {} & (original) & (original) & (GAU) & (GAU) & (BOX) & (BOX) \\
\noalign{\smallskip}\svhline\noalign{\smallskip}
96420-01-01-00 & I & 1.072 & NA & 0.675 & NA & 0.658 & NA \\
96420-01-11-00 & II & 1.366 & NA & 0.839 & NA & 1.212 & NA \\
96420-01-04-01 & III & 0.557 & NA & 0.699 & NA & 0.711 & NA \\
96420-01-05-00 & IV & 2.234 & NA & 3.623 & 6.485 & 1.705 & NA \\
96420-01-06-03 & V & 1.805 & NA & 3.315 & 4.492 & 2.165 & NA \\
96420-01-09-00 & VI & 0.928 & NA & 2.440 & NA & 3.587 & 6.330 \\
96420-01-18-05 & VII & 2.126 & NA & 8.388 & 3.015 & 1.388 & NA \\
96420-01-19-03 & VIII & 2.977 & NA & 4.950 & 4.731 & 5.217 & 4.731 \\
96420-01-35-02 & IX & 0.438 & NA & 2.823 & NA & 3.399 & 7.925 \\
\noalign{\smallskip}\hline\noalign{\smallskip}
\end{tabular}
\end{table}

\begin{table}[!h]
\caption{Correlation Integral results of IGR J17091-3624 (Part 2: ADA and NLM Data)}
\label{tab:CI2}       
%
\centering
\begin{tabular}{
  >{\centering\arraybackslash}p{2cm}  
  >{\centering\arraybackslash}p{1.2cm}  
  >{\centering\arraybackslash}p{1.2cm} 
  >{\centering\arraybackslash}p{1.2cm}  
  >{\centering\arraybackslash}p{1.2cm} 
  >{\centering\arraybackslash}p{1.2cm}
  >{\centering\arraybackslash}p{1.2cm}}
\hline\noalign{\smallskip}
ObsID & Class & $nmsd$ & $D_2$ & $nmsd$ & $D_2$ & Behaviour \\
{}& {} & (ADA) & (ADA) & (NLM) & (NLM) & {} \\
\noalign{\smallskip}\svhline\noalign{\smallskip}
96420-01-01-00 & I & 0.547 & NA & 1.160 & NA & S \\
96420-01-11-00 & II & 0.514 & NA & 0.807 & NA & S \\
96420-01-04-01 & III & 1.650 & NA & 1.011 & NA & S \\
96420-01-05-00 & IV & 2.410 & NA & 3.386 & 4.568 & S/NS \\
96420-01-06-03 & V & 5.204 & 3.247 & 0.596 & NA & S/NS \\
96420-01-09-00 & VI & 1.583 & NA & 4.347 & 5.242 & S/NS \\
96420-01-18-05 & VII & 5.486 & 3.190 & 4.149 & 3.243 & NS \\
96420-01-19-03 & VIII & 11.492 & 4.552 & 7.112 & 4.440 & NS \\
96420-01-35-02 & IX & 2.595 & NA & 6.444 & 6.043 & S/NS* \\
\noalign{\smallskip}\hline\noalign{\smallskip}
\end{tabular}
\end{table}

\begin{figure}[!htb]
    \centering
    \begin{minipage}[b]{0.45\linewidth}
        \includegraphics[width=1.2\linewidth, trim={1cm 0cm 1cm 0cm}, clip]{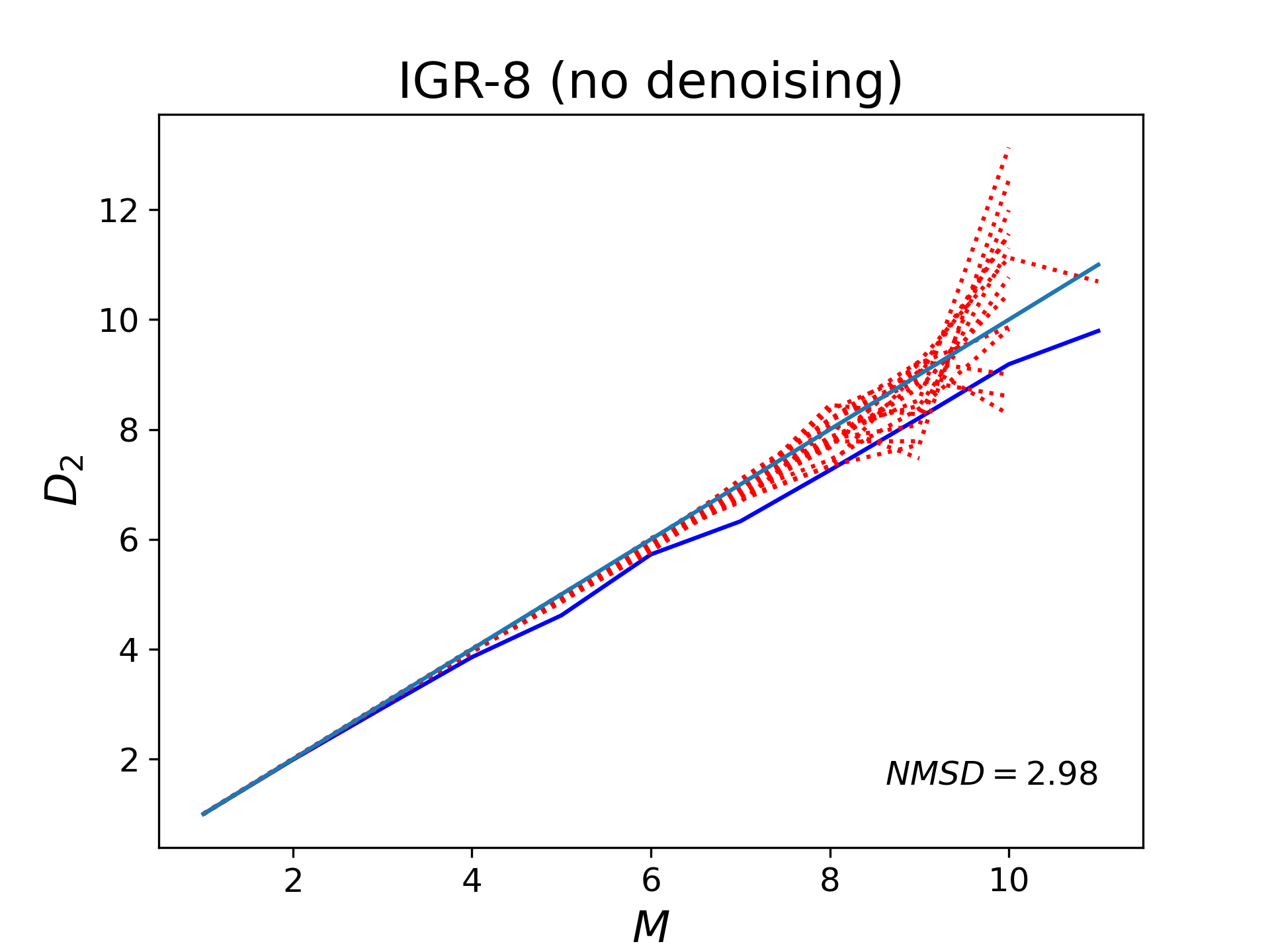}
    \end{minipage}
    \hfill 
    \begin{minipage}[b]{0.45\linewidth}
        \includegraphics[width=1.2\linewidth, trim={1cm 0cm 1cm 0cm}, clip]{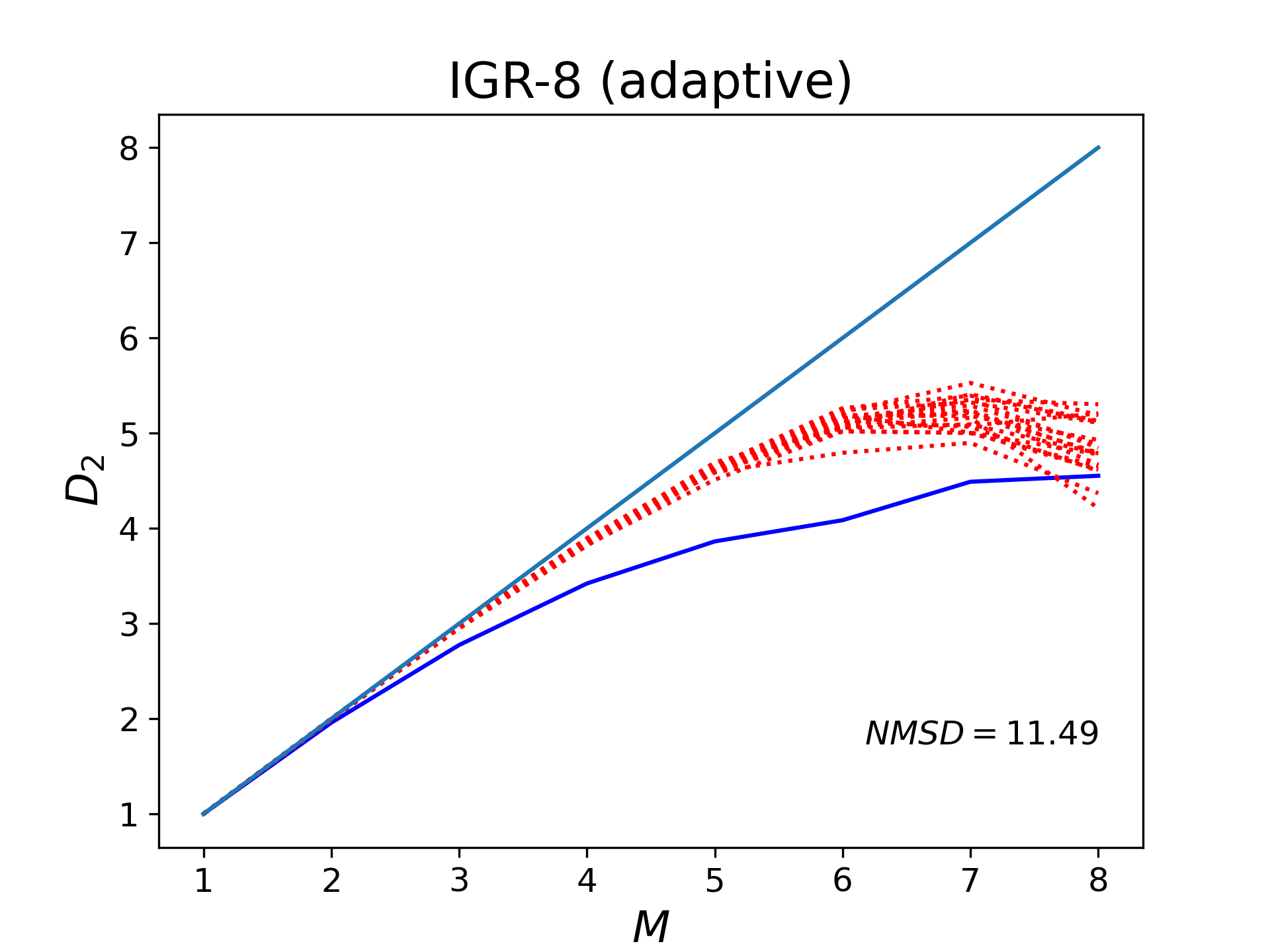}
    \end{minipage}

    \caption{The variation of $D_2$ with $M$ for the temporal class \rom{8} (determined to be S unanimously by all filtering techniques). The red dotted curves are the surrogates, and the dark blue solid curve is the true data. The light blue straight line shows $D_2 = M$, the expected result for an ideal S signal.}
    \label{fig: CD IGR-8}
\end{figure}

\begin{figure}[!htb]
    \centering
    \begin{minipage}[b]{0.45\linewidth}
        \includegraphics[width=1.2\linewidth, trim={1cm 0cm 1cm 0cm}, clip]{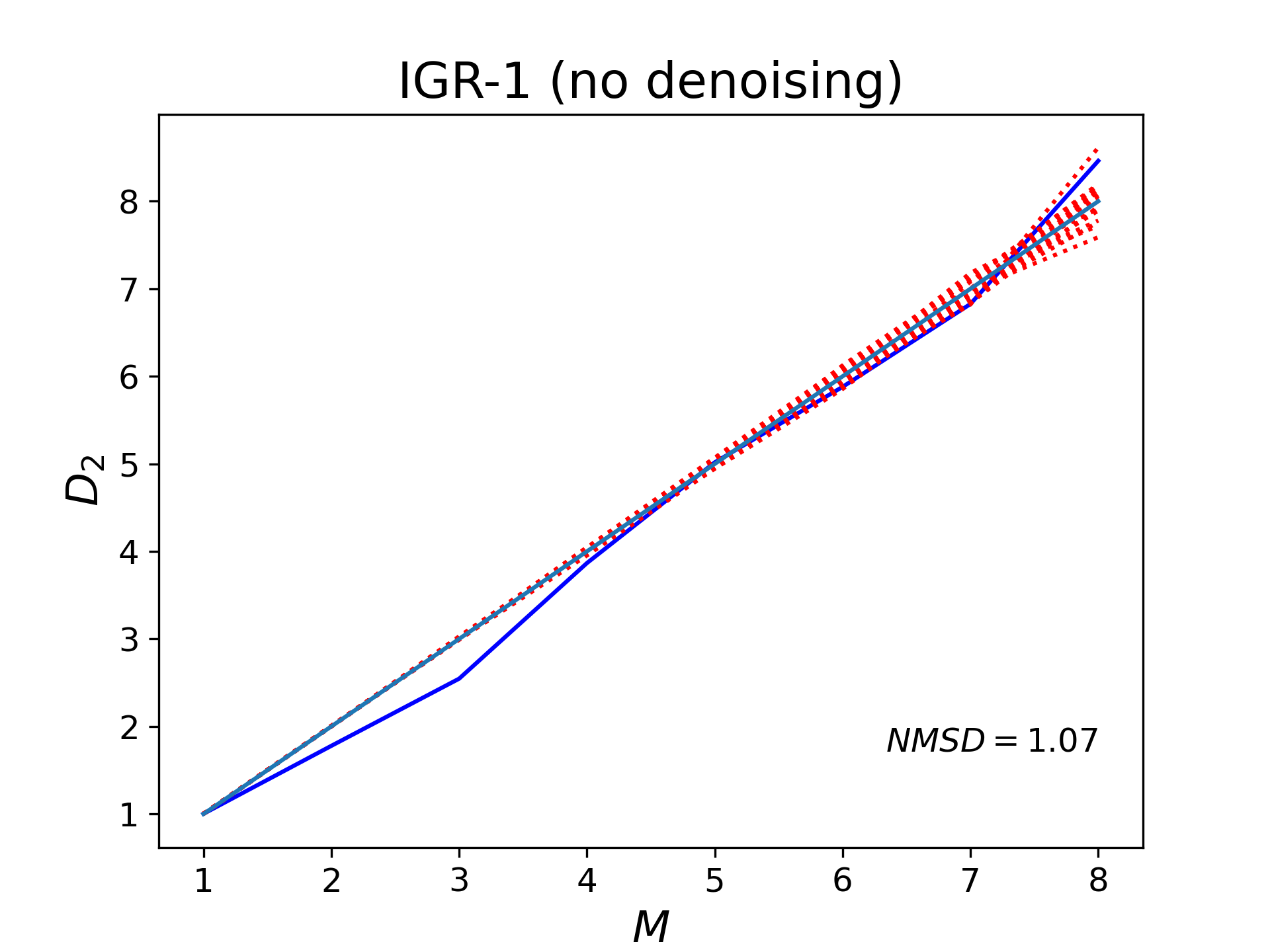}
    \end{minipage}
    \hfill 
    \begin{minipage}[b]{0.45\linewidth}
        \includegraphics[width=1.2\linewidth, trim={1cm 0cm 1cm 0cm}, clip]{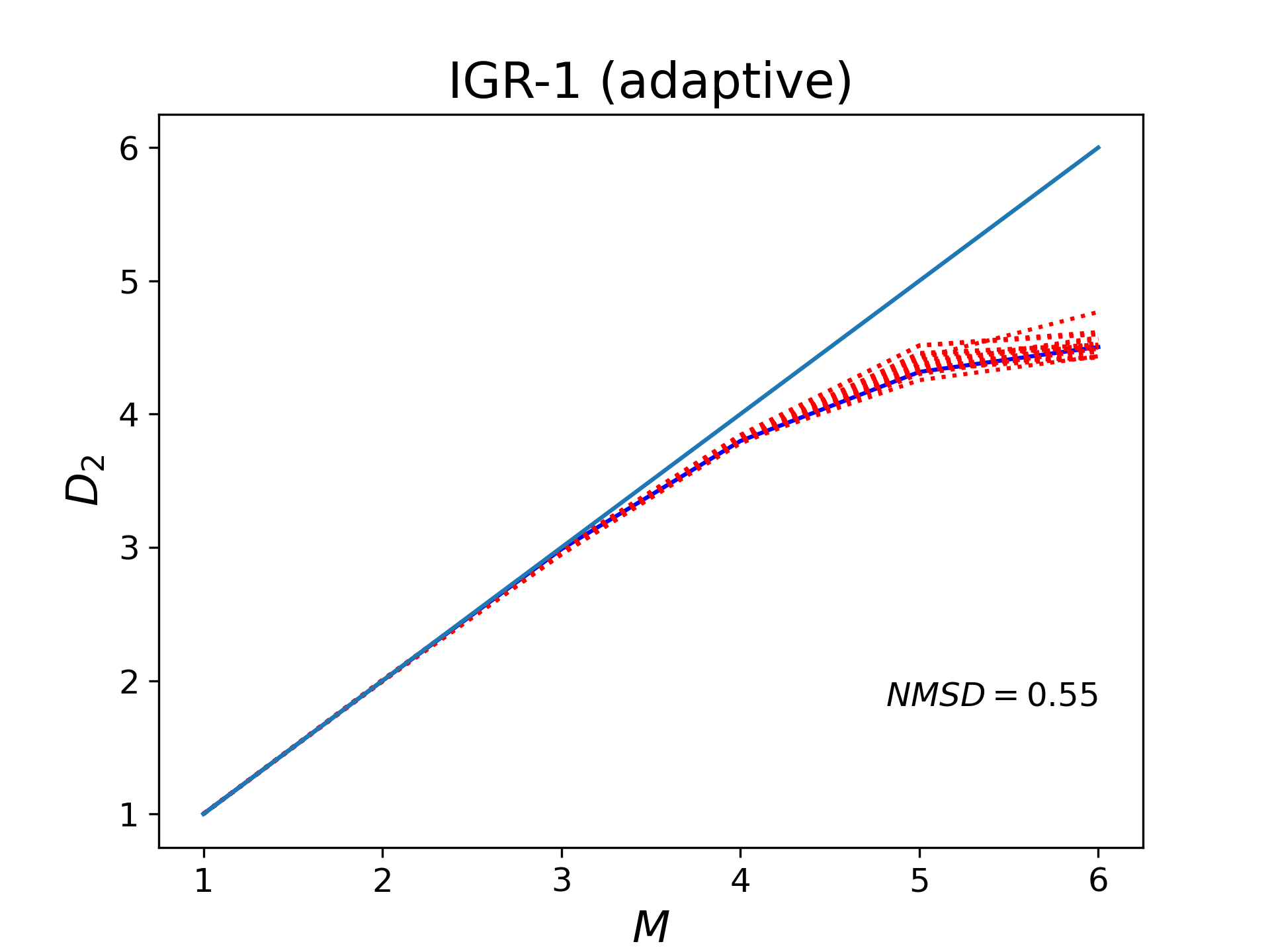}
    \end{minipage}

    \caption{The variation of $D_2$ with $M$ for the temporal class \rom{1} (determined to be S unanimously by all filtering techniques).}
    \label{fig: CD IGR-1}
\end{figure}

\subsection{Results from SVD} \label{subsec: SVD_result}

\begin{figure}[!htbp]
    \centering
    \begin{minipage}[b]{0.45\linewidth}
        \includegraphics[width=1.2\linewidth, trim={0cm 0cm 0cm 0cm}, clip]{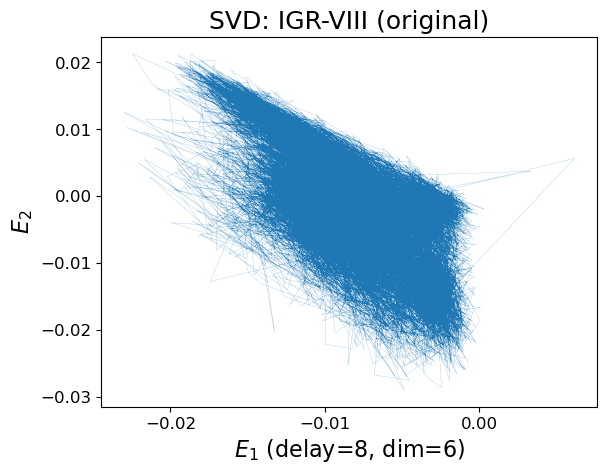}
    \end{minipage}
    \hfill 
    \begin{minipage}[b]{0.45\linewidth}
        \includegraphics[width=1.2\linewidth, trim={0cm 0cm 0cm 0cm}, clip]{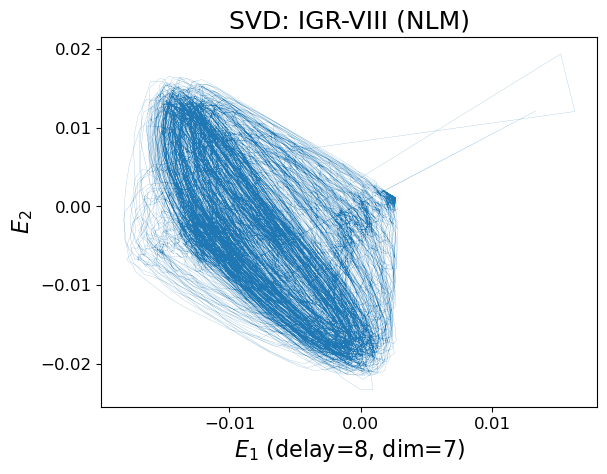}
    \end{minipage}

    \caption{SVD of IGR J17091-3624-\rom{8} without filtering and with NLM. The optimal embedding dimensions and delay are mentioned for each of them.}
    \label{fig: SVD IGR-8}
\end{figure}

\begin{figure}[!htbp]
    \centering
    \begin{minipage}[b]{0.45\linewidth}
        \includegraphics[width=1.2\linewidth, trim={0cm 0cm 0cm 0cm}, clip]{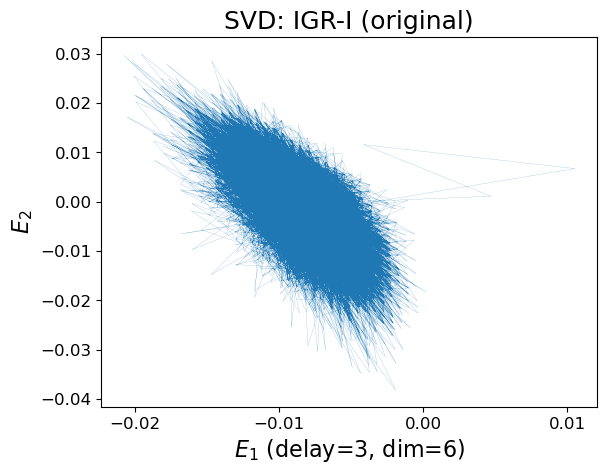}
    \end{minipage}
    \hfill 
    \begin{minipage}[b]{0.45\linewidth}
        \includegraphics[width=1.2\linewidth, trim={0cm 0cm 0cm 0cm}, clip]{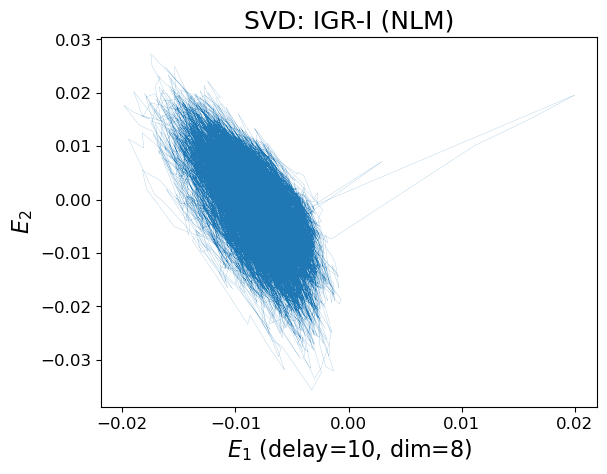}
    \end{minipage}

    \caption{SVD of IGR J17091-3624-\rom{1} without filtering and with NLM. It shows no complex structure post-filtering.}
    \label{fig: SVD IGR-1}
\end{figure}

SVD analysis reveals that classes \rom{5}, \rom{6}, \rom{7}, and \rom{8} exhibit complex trajectories with multiple loops and dense regions indicative of attractor behavior, reflecting high values of both $\beta_1$ and $\beta_0$. In contrast, classes \rom{1}, \rom{2}, \rom{3}, \rom{4}, and \rom{9} display simple topologies expected of S signals, specifically $(\beta_0, \beta_1) = (1,0)$. We have listed our results in Table \ref{tab:SVD}. Figures \ref{fig: SVD IGR-8} and \ref{fig: SVD IGR-1} contrast the complex phase portraits of class \rom{8} with the simpler blob-like structure of the S-type IGR J17091-362-\rom{1}, highlighting the effectiveness of our denoising techniques.

\begin{table}[!h]
\caption{Singular Value Decomposition results of IGR J17091-3624}
\label{tab:SVD}       
%
%
\centering
\begin{tabular}{
  >{\centering\arraybackslash}p{2.5cm}
  >{\centering\arraybackslash}p{1.2cm}
  >{\centering\arraybackslash}p{1.2cm}
  >{\centering\arraybackslash}p{1.2cm}
  >{\centering\arraybackslash}p{1.2cm}
  >{\centering\arraybackslash}p{1.2cm}
  >{\centering\arraybackslash}p{1.2cm}
  >{\centering\arraybackslash}p{1.2cm}}
\hline\noalign{\smallskip}
ObsID & Class & Original & ADA & NLM & GAU & BOX & Overall \\
\noalign{\smallskip}\svhline\noalign{\smallskip}
96420-01-01-00 & I & S & S & S & S & S & S \\
96420-01-11-00 & II & S & S & S & S & S & S \\
96420-01-04-01 & III & S & S & S & S & S & S \\
96420-01-05-00 & IV & S & NS & S & S & S & S \\
96420-01-06-03 & V & S & NS & NS & NS & NS & NS \\
96420-01-09-00 & VI & S & NS & NS & NS & NS & NS \\
96420-01-18-05 & VII & S & NS & NS & NS & NS & NS \\
96420-01-19-03 & VIII & S & NS & NS & NS & NS & NS \\
96420-01-35-02 & IX & S & NS & S & S & S & S \\
\noalign{\smallskip}\hline\noalign{\smallskip}
\end{tabular}
\end{table}

\subsection{Autoencoder Results} \label{subsec: Autoencoder results}
The used autoencoders are the same as in \cite{Pradeep2023} (code obtained from GitHub\footnote{https://github.com/csai-arc/blackhole\_stochasticity\_measure}). As mentioned in subsection \ref{subsec:Autoencoder}, the cutoff for classifying a signal as NS is $1.5$, which the authors have determined to provide optimum classification with test data. Our results are listed in Table \ref{tab:Autoencoder}.
\begin{table}[!h]
\caption{Deviation from stochasticity ($DS$) metric for IGR J17091-3624}
\label{tab:Autoencoder}       
%
%
\centering
\begin{tabular}{
  >{\centering\arraybackslash}p{2cm}  
  >{\centering\arraybackslash}p{1.2cm}  
  >{\centering\arraybackslash}p{1.2cm} 
  >{\centering\arraybackslash}p{1.2cm}  
  >{\centering\arraybackslash}p{1.2cm} 
  >{\centering\arraybackslash}p{1.2cm}
  >{\centering\arraybackslash}p{1.2cm}
  >{\centering\arraybackslash}p{1.2cm}} 
\hline\noalign{\smallskip}
ObsID & Class & $DS_{org}$ & $DS_{ADA}$ & $DS_{NLM}$ & $DS_{GAU}$ & $DS_{BOX}$ & Behaviour \\
\noalign{\smallskip}\svhline\noalign{\smallskip}
96420-01-01-00 & I & 0.13 & 0.79 & 0.78 & 0.53 & 0.52 & S \\
96420-01-11-00 & II & 0.11 & 0.66 & 0.63 & 0.46 & 0.46 & S \\
96420-01-04-01 & III & 1.04 & \textbf{4.80} & \textbf{5.55} & \textbf{3.79} & \textbf{3.81} & NS \\
96420-01-05-00 & IV & 1.27 & \textbf{2.54} & \textbf{3.06} & \textbf{2.28} & \textbf{2.30} & NS \\
96420-01-06-03 & V & 1.16 & \textbf{1.93} & \textbf{2.06} & \textbf{1.77} & \textbf{1.79} & NS \\
96420-01-09-00 & VI & 1.12 & \textbf{2.11} & \textbf{3.21} & \textbf{2.15} & \textbf{2.17} & NS \\
96420-01-18-05 & VII & 0.98 & 1.23 & 1.40 & 1.21 & 1.22 & S \\
96420-01-19-03 & VIII & 0.71 & 0.85 & 1.31 & 0.98 & 0.99 & S \\
96420-01-35-02 & IX & 0.69 & 0.84 & \textbf{2.55} & 1.26 & 1.27 & S \\
\noalign{\smallskip}\hline\noalign{\smallskip}
\end{tabular}
\end{table}
\subsection{PCA and DBSCAN Results} \label{subsec: PCA result}

\begin{figure}[htbp]
    \centering
    \begin{minipage}[b]{0.45\linewidth}
       \includegraphics[width=1.2\linewidth, trim={0cm 0cm 0cm 0cm}, clip]{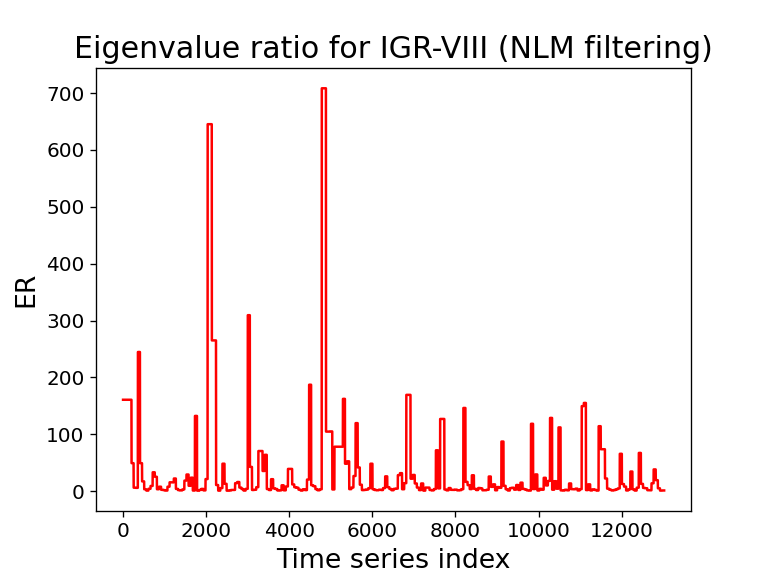} 
    \end{minipage}
    \hfill 
    \begin{minipage}[b]{0.45\linewidth}
         \includegraphics[width=1.2\linewidth, trim={3cm 0cm 1cm 0cm}, clip]{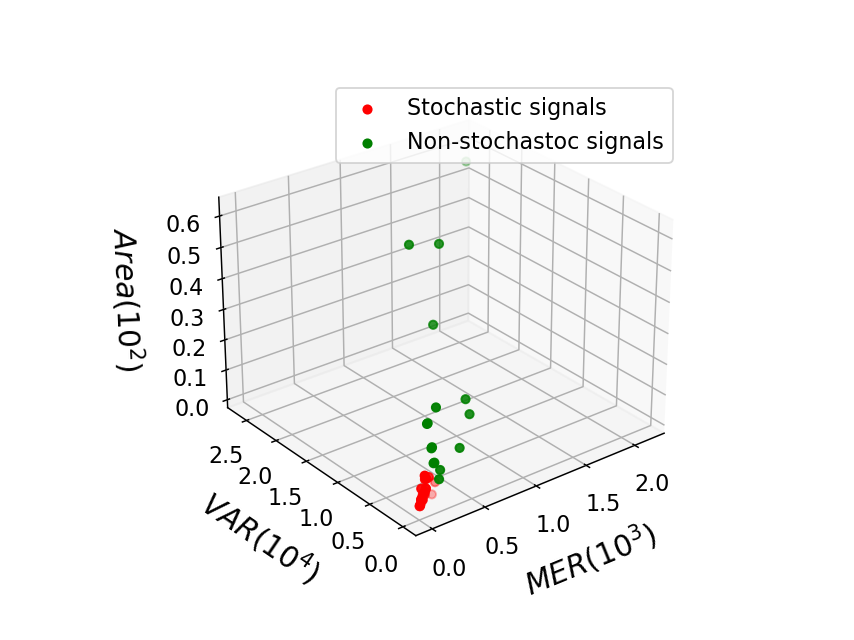}
    \end{minipage}

    \caption{\textbf{\textit{Left}:} A typical eigenvalue ratio (ER) plot. \textbf{\textit{Right}:} Scatter plot of PCA parameters from all (un)filtered signals.}
    \label{fig: classification}
\end{figure}

As outlined in subsection \ref{subsec: PCA}, the PCA algorithm has been applied to the original and filtered timeseries set, and the following results are obtained. The optimal eigenvalue ratio cutoff and $\epsilon$ for DBSCAN are chosen by maximizing the silhouette score \citep{Rousseeuw1987}. Figure \ref{fig: classification} shows the actual scatter of the cluster and the outliers. The red points are the S timeseries, while the green points are the outliers, which we identify here as the NS timeseries. Table \ref{tab:PCA_Summary} summarizes the results.

\begin{table}[!h]
\caption{Summary of PCA classification from different denoising methods}
\label{tab:PCA_Summary}
\centering
\begin{tabular}{
  >{\centering\arraybackslash}p{2cm}  
  >{\centering\arraybackslash}p{1cm}  
  >{\centering\arraybackslash}p{1.2cm}  
  >{\centering\arraybackslash}p{1.2cm}  
  >{\centering\arraybackslash}p{1.2cm}  
  >{\centering\arraybackslash}p{1.2cm}  
  >{\centering\arraybackslash}p{1.2cm}}  

\hline\noalign{\smallskip}
ObsID & Class & Original & ADA & NLM & GAU & BOX \\
\noalign{\smallskip}\hline\noalign{\smallskip}
96420-01-01-00 & I     & S & S & S & S & S \\
96420-01-11-00 & II    & S & NS & S & S & S \\
96420-01-04-01 & III   & S & S & S & S & S \\
96420-01-05-00 & IV    & S & NS & NS & NS & S \\
96420-01-06-03 & V     & S & NS & NS & NS & NS \\
96420-01-09-00 & VI    & S & NS & S & S & S \\
96420-01-18-05 & VII   & S & NS & NS & NS & NS \\
96420-01-19-03 & VIII  & S & S & NS & NS & NS \\
96420-01-35-02 & IX    & S & S & S & S & S \\
\noalign{\smallskip}\hline
\end{tabular}
\end{table}

\begin{table}[!h]
\centering
\caption{Summary of S/NS behavior across different methods}
\label{tab:summary}
\begin{tabular}{
  >{\centering\arraybackslash}p{2cm}  
  >{\centering\arraybackslash}p{1cm}  
  >{\centering\arraybackslash}p{1.2cm}  
  >{\centering\arraybackslash}p{1.2cm}  
  >{\centering\arraybackslash}p{1.4cm}  
  >{\centering\arraybackslash}p{1.2cm}
  >{\centering\arraybackslash}p{1.4cm}} 
\hline\noalign{\smallskip}
ObsID & Class & CI & SVD & Autoencoder & PCA & Overall \\
\noalign{\smallskip}\hline\noalign{\smallskip}
96420-01-01-00 & I & S & S & S & S & S\\
96420-01-11-00 & II & S & S & S & S & S\\
96420-01-04-01 & III & S & S & NS & S & S\\
96420-01-05-00 & IV & S/NS & S & NS & NS & S/NS\\
96420-01-06-03 & V & S/NS & NS & NS & NS & NS\\
96420-01-09-00 & VI & S/NS & NS & NS & S & S/NS\\
96420-01-18-05 & VII & NS & NS & S & NS & NS\\
96420-01-19-03 & VIII & NS & NS & S & NS & NS\\
96420-01-35-02 & IX & S/NS* & S & S & S & S\\
\noalign{\smallskip}\hline
\end{tabular}
\end{table}
\section{Summary}\label{sec:Discussion}

We have addressed Poisson noise contamination in X-ray lightcurves of IGR J17091-3624 using four filtering techniques: ADA, NLM, GAU, and BOX. These methods have facilitated the application of PCA, autoencoders, CI, and SVD to detect NS behavior in the filtered signals. We have summarized our findings in Table \ref{tab:summary} (we conclude upon S/NS division for a temporal class when three or more of our applied methods agree). IGR J17091-3624 classes \rom{1}, \rom{2}, and \rom{9} are consistently found to be S across all methods. Class \rom{3} is also predominantly S. In contrast, classes \rom{5}, \rom{7}, and \rom{8} are primarily NS. The classes \rom{4} and \rom{6} remain uncertain in their nonlinear features. Thus, we have uncovered potential complex dynamics in IGR J17091–3624 using multiple denoising methods.

Therefore, like GRS 1915+105, which sometimes exhibits NS or fractal (it could be deterministic and even chaotic) and sometimes S, IGR J17091-3624 also shows transient behavior switching between NS and S in various temporal classes. Such nature was unclear in the previous analysis \cite{Adegoke2020}, which could not denoise the signal appropriately. Nevertheless, earlier, both sources were classified into several temporal classes based on lightcurves and related combined temporal and spectral analyses, arguing their transient nature. The present study confirms a similar transient nature for them from the nonlinear dynamics point of view. This supports the argument of their twin nature. Many temporal classes of IGR J17091-3624 show the same dynamical behavior, e.g., S or NS, as their corresponding equivalent GRS 1915+105 classes, e.g., I \& $\chi$ and V \& $\mu$. However, one of the pairs of classes, i.e., III \& $\nu$, exhibits different dynamical behavior. The future plan is to combine the spectral behavior with the timeseries properties to understand the true nature of the underlying accretion flows, as was done earlier for GRS 1915+105 \cite{Adegoke2018}. The details will be reported elsewhere (Guria \& Mukhopadhyay, in preparation).

\bibliographystyle{spphys}
\bibliography{biblio}

\end{document}